# Molecular-beam epitaxy of monolayer and bilayer WSe$_2$: A scanning tunneling microscopy/spectroscopy study and deduction of exciton binding energy

H J Liu[1], L Jiao[1], L Xie[1], F Yang[2,3], J L Chen[1], W K Ho[1], C L Gao[2,3], J F Jia[2,3], X D Cui[1] and M H Xie[1*]

[1] Physics Department, The University of Hong Kong, Pokfulam Road, Hong Kong

[2] Key Laboratory of Artificial Structures and Quantum Control (Ministry of Education), Department of Physics and Astronomy, Shanghai Jiaotong University, 800 Dongchuan Road, Shanghai 200240, China

[3] Collaborative Innovation Center of Advanced Microstructures, Department of Physics and Astronomy, Shanghai Jiao Tong University, Shanghai 200240, P. R. China

Email: mhxie@hku.hk

**Abstract.** Interests in two-dimensional (2D) transition-metal dichalcogenides (TMDs) have prompted some recent efforts to grow ultrathin layers of these materials epitaxially using molecular-beam epitaxy (MBE). However, growths of monolayer (ML) and bilayer (BL) WSe$_2$ – an important member of the TMD family – by the MBE method remain uncharted probably because of the difficulty in generating tungsten fluxes from the elemental source. In this work, we present a scanning tunneling microscopy and spectroscopy (STM/S) study of MBE-grown WSe$_2$ ML and BL, showing atomically flat epifilm with no domain boundary (DB) defect. This contrasts epitaxial MoSe$_2$ films grown by the same method, where a dense network of the DB defects is present. The STS measurements of ML and BL WSe$_2$ domains of the same sample reveal not only the bandgap narrowing upon increasing the film thickness from ML to BL, but also a band-bending effect across the boundary(step) between ML and BL domains. This band-bending appears to be dictated by the edge states at steps of the BL islands. Finally, comparison is made between the STS-measured electronic bandgaps with the exciton emission energies measured by photoluminescence, and the exciton binding energies in ML and BL $WSe_2$ (and $MoSe_2$) are thus estimated.

1. Introduction

Two-dimensional (2D) transition-metal dichalcogenides (TMDs) have sizable energy bandgaps, strong spin-orbit coupling and valley-contrasted properties. They offer new platforms for exploring low-dimensional physics and promise ultrathin or 2D electronics, optoelectronics, and the emerging spin- and valley-tronic applications [1]. Soon after the first successful isolation by exfoliation of $MoS_2$ monolayer (ML, defined hereafter as the X-M-X trilayer, where X stands for the chalcogen atom and M refers to the metal) [2], fabrications of ultrathin TMD films by some more controlled methods, such as hydrothermal synthesis [3], chemical vapor deposition (CVD) [4, 5] and molecular-beam epitaxy (MBE) [6-10], have been attempted. With the advantage of thickness and doping controls and the readily available surface characterization tools *in situ*, the MBE method has drawn increasing attention for the growth of ultrathin TMD layers for some surface and electronic studies. Indeed, MBE growths of ML and bilayer (BL, i.e., two X-M-X trilayers) $MoSe_2$ have been reported by a few groups [6-10]. In contrast, MBE growth of $WSe_2$, another important member of the TMD family, remains uncharted. Intriguingly, the as-grown ultrathin $MoSe_2$ films have been shown to contain dense networks of inversion domain boundary (DB) defects, which give rise to mid-gap states with electronic, optical and catalytic consequences. Therefore, it will be of great fundamental and application interests to examine the characteristics of the MBE-grown TMD films other than $MoSe_2$. In this work, we demonstrate MBE-grown $WSe_2$ epifilms free from the DB network. Background doping in as-grown film is low, indicating better quality than as-grown $MoSe_2$ samples. Differential conductance (dI/dV) spectra taken by scanning tunneling spectroscopy (STS) measurements of both ML and BL WSe$_2$ reveal not only bandgap narrowing upon film thickness increase from ML to BL, but also a band-bending effect towards the boundaries(steps) between ML and BL domains of the same sample. Finally, by comparing the STS-measured electronic energy bandgaps with exciton emission energies in photoluminescence (PL) spectra, exciton binding energies of ML and BL WSe$_2$ are deduced. For completeness, results of ML and BL MoSe$_2$ are also presented.

2. **Experimental**

Growths of WSe$_2$ (and MoSe$_2$) ultrathin films were carried out in a customized Omicron MBE system with the base pressure of in the low $10^{-10}$ mbar range. Elemental W and Mo metal wires were used as the metal sources in the EFM-3 e-beam evaporators (without ion filtering) from Omicron NanoTechnology GmbH, while elemental Se source in a dual-filament Knudsen cell was heated to 120 $^o$C with the "hot-lip" temperature set at 220 $^o$C in order to prevent Se condensation at the cell orifice. The fluxes of the metal sources were calibrated by the built-in flux monitors in the e-beam evaporators, and that of Se was estimated by the beam-equivalent pressure (BEP) measured using a beam flux monitor at the sample position. During film deposition, the BEP of Se was about $1.1 \times 10^{-6}$ mbar while the background pressure in the chamber was ~ $1.2 \times 10^{-8}$ mbar. The metal to Se flux ratio was 1 : 15 and the deposition rate was 0.5 MLs/hr according to post-growth coverage/thickness measurements of the deposits. Freshly cleaved highly ordered pyrolytic graphite (HOPG) substrate was degassed in the ultrahigh vacuum (UHV) chamber for overnight and flashed at 550 $^o$C before commencing the MBE experiment at 300 – 450 $^o$C. The substrate temperature was estimated by the power given to the heating filaments (W) in the sample manipulator, which had been calibrated by the melting points of three high purity (> 4N) elements of indium (156.60 $^o$C), selenium (217 $^o$C) and bismuth (271.5 $^o$C) [11] placed on the surface of a HOPG wafer. During deposition, the sample surfaces were monitored in situ by reflection high-energy electron diffraction (RHEED) operated at 10 keV. The observation of the streaky RHEED patterns indicated the layer-by-layer growth mode of WSe$_2$ (and MoSe$_2$) on HOPG. After a preset coverage of the film was deposited, the source fluxes were stopped by closing the mechanical shutters in front of the source cells and in the meantime, the sample was cooled naturally to room temperature (RT) for subsequent scanning tunneling microscopy (STM) experiments at RT in an adjacent UHV chamber using an Omicron VTSTM facility. Afterwards, the sample was transferred back into the MBE reactor for deposition at RT of an amorphous Se "capping" layer. It was then taken out of the UHV and transported to a separate Unisoku low-temperature (LT) STM system for LT-STM/S measurements at 4 K or 77 K. Prior to the LT-STM/S experiments, however, the Se capping layer was thermally desorbed at ~ 200 $^oC$ for half an hour, which was confirmed by the recovery of the sharp and streaky RHEED patterns as well as the revelation of the same step-and-terrace morphology of the surface by STM. For all STM/S measurements, the constant current mode was adopted. In addition, the STS measurements were performed using the lock-in technique with the modulation voltage 15 mV and frequency 985 Hz. Each presented STS curve in the following (and in Supplementary) represents an average of 50 measurements at the same location of the sample and for the same constant tip-sample distance.

PL experiments of ML and BL $WSe_2$ and $MoSe_2$ were carried out with a confocal-like setup with an excitation source of 532 nm at temperatures 10-300K. Instead of the MBE-grown samples, atomically thin flakes exfoliated from bulk single crystals on 300nm-$SiO_2$ capped Si wafers were used for the PL experiments. This was because the photoluminescence from the MBE films was fully quenched by the HOPG substrate. Although non-ideal, comparison between the PL emission peak energies and the STS-measured electronic bandgaps would allow us to perform order-of-magnitude estimates of exciton binding energies in both ML and BL $WSe_2$ and $MoSe_2$.

**3. Results and discussions**

*3.1 STM/S of ML and BL $WSe_2$ and a comparison with $MoSe_2$*

Figure 1(a) presents a STM image of an as-grown $WSe_2$ film of nominal thickness of 1.2 MLs. The film is predominantly ML $WSe_2$, but there is also appreciable coverage of BL domains/islands and holes of exposed substrate due to the kinetics of MBE process. The terrace-and-step morphology of the surface and the streaky RHEED pattern (see inset) suggest the layer-by-layer growth mode of WSe$_2$ on HOPG, which resembles that of MoSe$_2$ growth on the same substrate [12]. However, we note a striking difference between epitaxial WSe$_2$ and MoSe$_2$. As reported in an early publication, the MBE-grown MoSe$_2$ film often contains high density of the DB defects intertwined into a triangular network [7]. These DB defects manifest in the STM micrographs as bright lines when imaged at the bias conditions corresponding to the gap region of MoSe$_2$ (see figure 3(a) below). In epitaxial WSe$_2$, on the other hand, no such bright line is found. In the atomic resolution STM image of figure 1(b), one only observes regular moiré patterns due to the lattice misfit between WSe$_2$ and graphite but no sign of the DB defect. Given the same crystal structure and similar lattice parameters between WSe$_2$ and MoSe$_2$, it is somewhat surprising that the two materials behave so differently during MBE growth on HOPG. The WSe$_2$ film is thus more attractive and advantageous for studying the intrinsic properties of ultrathin TMDs.

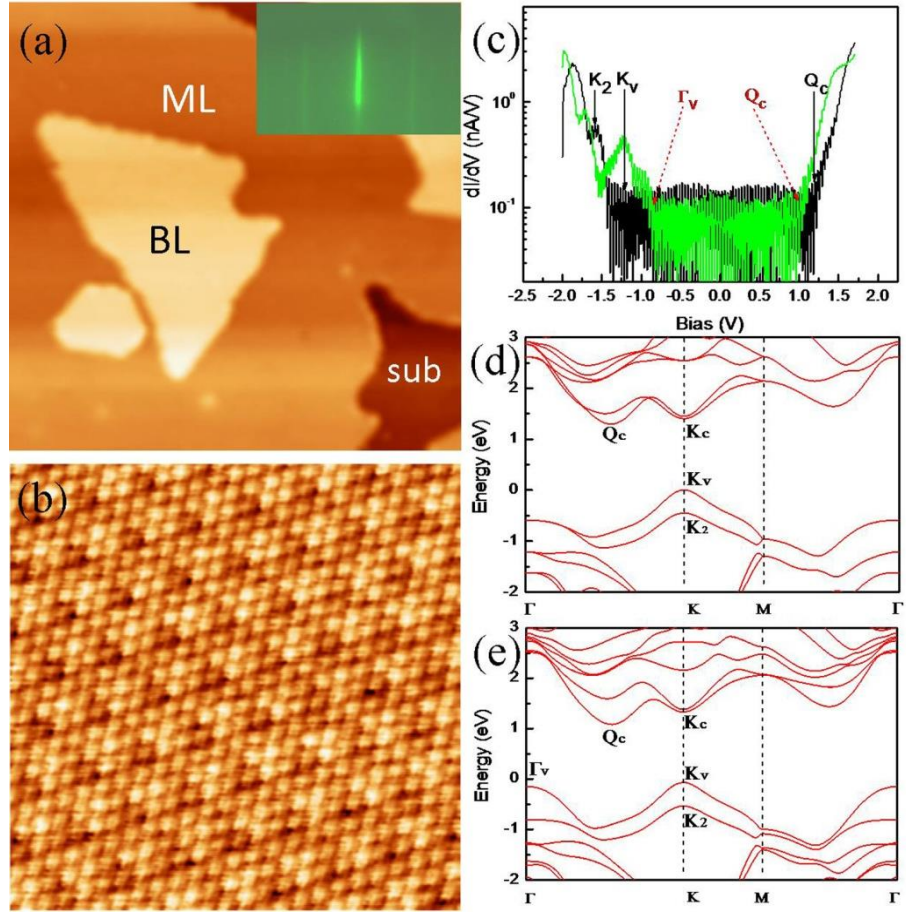

**Figure 1.** STM/S of MBE-Grown ML and BL WSe$_2$. (a) STM micrograph (size: 75×75 nm$^2$, sample bias: 2.4 V) of a MBE-grown WSe$_2$ film with the nominal thickness of 1.2 MLs, showing ML and BL domains. Holes of exposed substrate surface are also visible. The inset shows the RHEED pattern taken along [11$\bar{2}$0] of the surface. (b) A close-up, atomic resolution STM image (size: 7.5×7.5 nm$^2$, sample bias: 0.8 V) of the ML WSe$_2$ domain of the sample revealing the moiré pattern but no line defect. (c) STS differential conductance spectra of both ML (black) and BL (green) WSe$_2$, each represents an average of 50 measurements at fixed locations on sample and the same tip-sample distance. The critical point energies are indicated by solid (for ML) and dashed (for BL) arrows. (d, e) Theoretical band structures of ML and BL WSe$_2$, respectively, calculated by the DFT, in which the critical points are labelled.

In figure 1(c), we present two differential conductance spectra obtained by the STS measurements on ML and BL regions of the same sample and at fixed positions far from steps. Both curves reveal semiconductor property with sizable energy gaps. The Fermi level (0 eV) is found close to the middle of the energy gaps, indicative of low background doping of the film. This also contrasts the MoSe$_2$ film grown by MBE, which is often electron doped and the Fermi level is close to the conduction band edge (refer to figure 3(b)). Comparing the two spectra in figure 1(c), one clearly notes a gap narrowing effect upon film thickness increase from ML to BL.

We follow the method of Ref. [8, 10] (see Supplementary Materials) to locate the band edges from the dI/dV spectra of figure 1(c) and thus determine the energy bandgaps of ML and BL WSe$_2$. They are 2.59±0.07 eV and 1.83±0.10 eV for ML and BL WSe$_2$, respectively. The value of 2.59 eV for ML WSe$_2$ is consistent with an early reported result [13], but as suggested in a recent study, the experimental spectral edge may not reflect the true electronic band edges of ML WSe$_2,$ and consequently the "apparent" energy gap is usually overestimated in the STS spectrum by as much as the spin-split $\Delta_{SO}$ at the K point of the Brillouin zone (BZ) [8, 14]. For ML WSe$_2$, the $\Delta_{SO}$ is as high as 0.4 eV [8, 14]. Figure 1(d) and 1(e) present the calculated energy bands by the density functional theory (DFT) for ML and BL WSe$_2$, respectively, taking account the spin-orbit coupling. The fact that the valence states at K are mainly of the $d_{x^2-y^2}$ and $d_{xy}$ components of the metal atoms and they have large in-plane momenta ($k_\parallel$) make them less sensitive to STS measurements [8, 15]. The experimental spectral edge below Fermi energy does not necessarily reflect the $K_v$ state of the valence band maximum (VBM) (see figure 1(d)) [8, 15], and according to Zhang *et al*. [8], the first peak close to the spectral edge of the valance band coincides with the lower spin-split band $K_2$, which is at -1.61 eV in figure 1(c) (black). Given the 0.4 eV spin-splitting, the VBM is thus more likely located at -1.21 eV as marked in the figure. For the conduction band, there is an ambiguity where the conduction band minimum (CBM) lies – the K or the Q valley – and the latter is approximately midway between Γ and K (refer to figure 1(d)) [14]. Recent STS experiment [8] suggested the Q-valley to be 0.08 eV below $K_c$. In any case, as the $Q_c$-state contains much of the *p*-orbital component of the chalcogen atoms and has a lower $k_\parallel$ than the $K_c$-state, the STS spectral edge above the Fermi energy likely corresponds to $Q_c$-state as marked and it is found to be at +1.18 eV. Therefore, we derive the indirect bandgap (i.e., $Q_c$ – $K_v$) of ML WSe$_2$ to be about 2.39 eV.

For BL WSe$_2$, the CBM is affirmed at $Q_c$, which can be found at +0.99 eV in figure 1(c) (green). For the VBM, the energy difference between $\Gamma_v$ and $K_v$ (refer to figure 1(e)) is small. But as the STS measurement is more sensitive to the Γ-states, the spectral edge at -0.84 eV likely reflect the $\Gamma_v$-state and we derive an energy gap of BL WSe$_2$ to be 1.83±0.10 eV, which corresponds to the indirect gap between $Q_c$ and $\Gamma_v$.

Because the two spectra in figure 1(c) are from two close-by positions of the same sample but of different thickness domains, it is reasonable to assume the ML and BL share the same Fermi level. In figure 1(c), one notes the different magnitudes of the band-edge energy shifts of the valence versus conduction band edges: 0.37 eV for the VBM and 0.19 eV for the CBM, which give rise to an overall bandgap narrowing of 0.56 eV in BL WSe$_2$ over that of ML. While the 0.19 eV shift of the CBM is for the same $Q_c$ bands, which may reflect the inter-layer coupling of the *p*-orbital electrons in BL WSe$_2$, the 0.37 eV energy shift of the VBM may however be thought of the change of the VBM from the K valley in ML WSe$_2$ to Γ in BL film. On the other hand, we also note a band-bending effect at the boundary (i.e., step) of ML and BL domains. The latter may cause an additional shift of the band-edges as illustrated in figure 2(a) and discussed below.

Band-bending close to the boundary of ML and BL domains is evidenced by the STS

spectra taken at varying distances from a step as shown in figure 2(b) and 2(c) for the ML and BL regions, respectively. Upon approaching the step, the spectra show a blue-shift for both domains as illustrated in figure 2(a). The magnitudes of the upward shift (i.e., the band-bending) can be different in ML versus BL domains. This will result in an "apparent" electronic band misalignment even far from the step where the spectra in figure 1(c) were measured.

The upward band-bending at both sides of the junction is unusual and interesting. We attribute it to the Fermi level pinning by the in-gap states associated with the edge atoms at the step. It was demonstrated that edge atoms of ML $MoS_2$ clusters could induce in-gap states close to the VBM [16]. If the bulk of the film (i.e., away from the step) is nearly intrinsic, i.e., the Fermi-level is in the middle of the bandgap as suggested by figure 1(c), then the edge atoms at the junction will induce an upward band-bending, resulting in the band diagram of figure 2(a). This type of band-bending implies depletion of conduction electrons but accumulation of holes at the step, which may lead to some interesting electronic and optical effects.

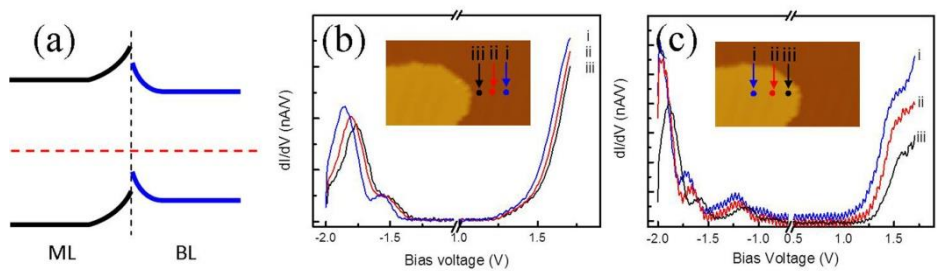

**Figure 2.** STS of BL $WSe_2$ at different locations. (a) Schematic diagram of electronic bands at the boundary between ML and BL $WSe_2$, showing the band-bending and band-edge shifting effect (relative to the Fermi level) due to different electron affinity of the two films. (b, c) STS differential conductance spectra of ML (b) and BL (c) $WSe_2$ at varying distances from the step edge (see insets), revealing the band-bending effect.

We now make a comparison with the STS results from ML and BL $MoSe_2$. Figure 3(a) shows an as-grown $MoSe_2$ film of 1.4 MLs nominal coverage. As pointed out earlier, such films contain networks of the DB defects in both ML and BL domains. The conductance spectra presented in figure 3(b) were taken from points away from both steps and the DB defects in order to reveal the intrinsic properties of MoSe$_2$ layers. Following the same procedure, we derive the energy gaps of ML and BL MoSe$_2$ to be 2.25±0.05 eV and 1.72±0.05eV, respectively, from the STS spectra. Unlike WSe$_2$, there is no ambiguity about the CBM for ML MoSe$_2$, which is in the K valley. The valence bands at K are spin-split by ~ 0.18 eV [17]. Given that the STS does not necessarily reveal the band edges at K, the measured gap of 2.25 eV may represent an upper bound of the true gap of ML MoSe$_2$. As for BL MoSe$_2$, the VBM is known to be shifted to $\Gamma_v$, while the CBM is at $Q_c$. Both are sensitive to STS measurements, so the STS-measured gap of 1.72 eV more likely reflects the true indirect gap ($Q_c$ - $\Gamma_v$) of BL MoSe$_2$. Finally, similar to WSe$_2$, both the conduction and valence band edges are seen shifted upon going from ML to BL of the sample.

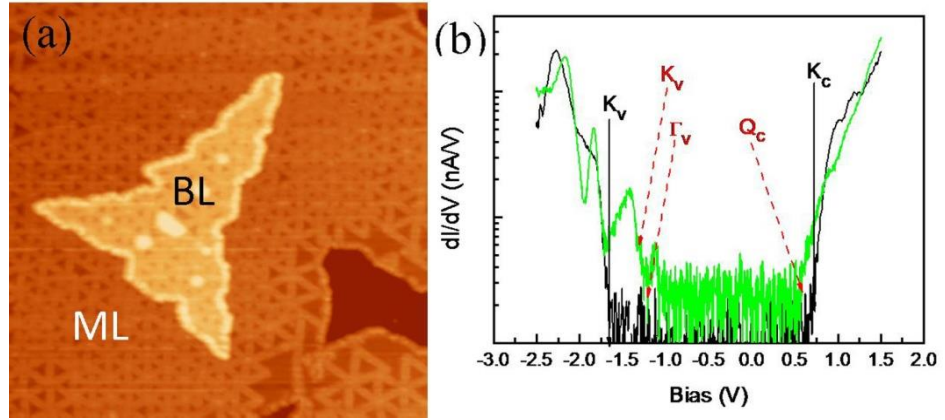

**Figure 3.** STM/S of ML and BL MoSe$_2$. (a) STM image (size: 100×100 nm$^2$, sample bias: -1.0 V) of an as-grown MoSe$_2$ film of the nominal thickness of 1.4 MLs, showing the network of domain boundary defects (the bright lines) in both ML (darker area) and BL (brighter area) domains. (b) STS differential conductance spectra of ML (black) and BL (green) MoSe$_2$.

*3.2 Exciton binding energies of ML and BL WSe$_2$ and MoSe$_2$*

Having determined the electronic bandgaps by the STS of ML and BL WSe$_2$ and MoSe$_2$ in the above, we estimate exciton binding energies in these materials by comparing with the PL results of the same materials. Indeed, giant excitonic effect in ultrathin TMDs originating from enhanced Coulomb interaction between electrons and holes due to spatial confinement and reduced dielectric screening is one of the remarkable properties of the 2D systems, which are attracting extensive theoretical and experimental attention lately [10,14,18-27]. The reported size of exciton binding energy, a key quantity describing the strength of excitonic effects, has not been very consistent, ranging from ~300 meV to about 1.0 eV for ML TMDs [10,14,18-27]. Despite such variations, they represent an order of magnitude increase over that in the bulk [28] and in conventional 2D semiconductor quantum wells [29].

It is unfortunate that the PL measurements of the same MBE samples studied by the STS above are not successful. No band-edge emission was detected and it was likely due to the quenching effect by the highly conductive HOPG substrate. We thus performed the PL experiments on exfoliated samples instead. Although non-ideal, the results still provide evidence of large exciton binding energies of the direct emissions in ML $WSe_2$ and $MoSe_2$ as well as reduced exciton binding energies in BL films for the indirect exciton emissions.

Figure 4(a) to 4(d) present the PL spectra for both $MoSe_2$ (a and b) and $WSe_2$ (c and d) ML and BL samples. Because the luminescence intensities of the ML samples are around one order of magnitude higher than the BL counterparts, we have shown the normalized PL intensities in all plots for clarity. Comparing figure 4(a) and 4(b), we find that the normalized PL spectra from $MoSe_2$ are almost identical for the ML and BL films in both the emission peak position and the intensity ratio. The only noticeable difference is the intensity shoulder seen in figure 4(b) around 1.6 eV, which can be attributed to the *indirect* exciton emission according to its temperature dependent behavior as elaborated in the Supplementary. The strong PL peaks at 1.657 eV and 1.629 eV are the *direct* exciton and trion emissions, respectively [17]. As ML $MoSe_2$ has an energy bandgap of 2.25 eV, one deduces that the exciton binding energy in ML $MoSe_2$ is about 0.59 eV, which likewise represents an upper bound due to the overestimate of the bandgap by STS. For BL $MoSe_2$, the STS-measured gap (1.72 eV) reflects the indirect gap between $Q_c$ and $\Gamma_v$, while the dominant exciton emission (1.657 eV) is from the direct emission at K valley. So a simple subtraction of the two energies is not very meaningful. As noted earlier, the weak intensity shoulder at about 1.6 eV in figure 4(b) reflects the indirect exciton emission. By multiple peak fitting of the PL spectrum, we locate the indirect exciton emission peak at 1.602 eV, which we attribute to $Q_c$-$K_v$ emission by consideration of the second-order Moller-Plesset perturbation [29]. The local minimum in the conduction band at Γ is much higher than that at K (cf. figure 1(e)). According to the Moller-Plesset perturbation theory [30], phonon-assisted indirect emission $Q_c$-$K_v$ would overwhelm that of $Q_c$-$Γ_v$. By noting the energy gap between $Q_c$ and $K_v$ is ~1.81 eV, one deduces that the binding energy of indirect exciton in BL MoSe$_2$ is about 0.21 eV.

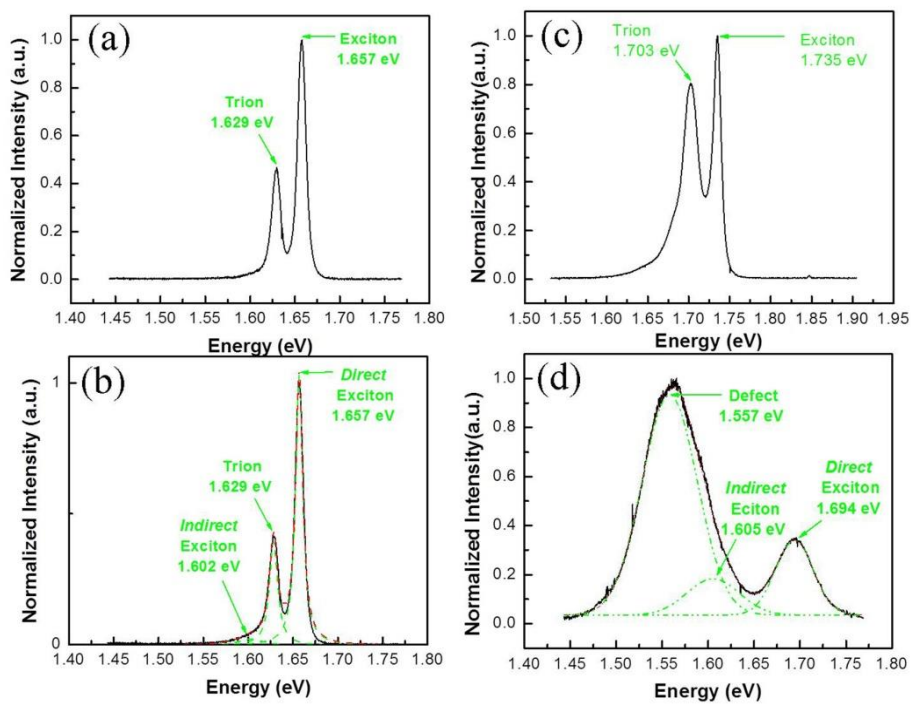

**Figure 4.** PL spectra of ML and BL MoSe$_2$ and WSe$_2$. (a) Normalized PL spectrum of ML MoSe$_2$ measured at 10 K. Exciton and trion emissions are identified. (b) Normalized PL spectrum of BL MoSe$_2$ measured at 10 K. Multiple-peak fitting (dashed green and red lines) resolves direct and indirect exciton emissions and that of trion. (c) Normalized PL spectrum of ML WSe$_2$ measured at 77K revealing exciton and trion emissions. (d) Normalized PL spectrum of BL WSe$_2$ measured at 77K. Multiple-peak fitting (dashed green and red lines) resolves direct and indirect exciton emissions besides a possible defect-related luminescence.

For WSe$_2$, our PL experiments have revealed complications, showing some unknown peaks when measured at 10K (see Supplementary). These peaks are likely related to defects or traps in sample, and their presence hinders accurate assignments of exciton emission energies. On the other hand, we have found the PL spectra are clearer at lifted temperatures. The data measured at 77 K are used for extracting the exciton binding energies in WSe$_2$ as shown in figure 4(c) and 4(d). For ML sample (figure 4(c)), the exciton and trion emissions are identified at 1.735 and 1.703 eV, respectively. By comparing with the STS-measured bandgap at the same temperature (2.39 eV), one deduces the exciton binding energy of 0.72 eV. For BL WSe$_2$, we do not observe trion emission but a broad peak at about 1.56 eV (figure 4(d)). The broad peak vanishes at high-temperatures (e.g., 180 K). So this peak is again likely related to defects. The PL evolution as a function of temperature (10-300K) and the multi-peak fitting of the spectrum lead to the direct and indirect exciton emission peaks at 1.694 and 1.605 eV at 77K, respectively. As we only know the indirect bandgap of BL WSe$_2$ from the STS measurements (1.83 eV), noting further that there is a small energy difference between $\Gamma_v$ and $K_v$, we deduce for the indirect exciton binding energy in BL WSe$_2$ to be about 0.23 eV. The reduced exciton binding energy in BL TMD is the result of indirect gap, where the electron and holes do not share the same crystal momentum. The significant difference of the exciton binding energy between ML and BL TMD thus reflects the band edge shift. Unfortunately due to the ambiguity of the direct gap (K valley) in the STS measurement, we cannot estimate the direct-gap exciton binding energy of BL.

Table 1 summarizes the deduced exciton binding energies for both ML and BL WSe$_2$ and MoSe$_2$. We note the binding energy of 0.59 eV for ML MoSe$_2$ is in reasonable agreement with that found in ML MoSe$_2$ grown on graphene/SiC (0.55 eV) [10], while that of ML WSe$_2$ (0.72 eV) appears consistent with the previously reported value of 0.79 eV in Ref. [24] and 0.6±0.2 eV in [25]. For BL TMD films, the exciton binding energies become smaller but they are of the indirect excition emissions as compared to the direct emission in ML films.

**Table 1.** Exciton binding energies in ML and BL WSe$_2$ and MoSe$_2$

|  | ML (direct) | BL (indirect) |
|---|---|---|
| WSe$_2$ | 0.72 ±0.07 eV | 0.23 ±0.10 eV |
| MoSe$_2$ | 0.59 ±0.05 eV | 0.21 ±0.05 eV |

## 4    Conclusions

To conclude, we have grown atomically flat WSe$_2$ films on HOPG by the method of MBE, in which the domain boundary network is absent. Differential conductance spectra taken from both ML and BL WSe$_2$ and MoSe$_2$ show not only semiconductor films and bandgap narrowing when changing the film thickness from ML to BL, but also shift of both conduction and valence band edges. Band-bending at the boundary (step) of ML and BL domains are evidenced, which is attributed to a Fermi-level pinning effect by states of the step-edge atoms. Energy bandgaps of the materials are derived and compared with the PL spectra from exfoliated samples. Exciton binding energies in ML and BL WSe$_2$ and MoSe$_2$ are estimated. The results provide another experimental evidence of the large excitonic effect in ultrathin TMDs as well as the apparent reduction of the exciton binding energy for the *indirect* emission in BL TMDs than that of the *direct* emission in ML samples.

**Acknowledgements**

We thank G.B Liu for proving the DFT calculated bands of WSe$_2$ and MoSe$_2$ and W. Yao for some insightful comments. We acknowledge the support of the CRF grant (No. HKU9/CRF/13G) from the Research Grant Council of Hong Kong Special Administrative Region, China. MHX and JFJ acknowledge the support from the MoE/RGC joint research grant (No. M-HKU709-12). HJL and MHX also acknowledge the support from the internal grants of The University of Hong Kong. The work in SJTU was supported by the MOST of China (2013CB921902, 012CB927401) and the NSFC (11227404, 11374206).

**References**

[1] Wang Q H, Kalantar-Zadeh K, Kis A, Coleman J N and Strano M S 2012 Electronics and optoelectronics of two-dimensional transition metal dichalcogenides *Nat. Nano.* **7** 699-712

[2] Novoselov K S, Jiang D, Schedin F, Booth T J, Khotkevich V V, Morozov S V and Geim A K 2005 Two-dimensional atomic crystals *Proc. Natl. Acad. Sci. USA* **102** 10451-3

[3] Peng Y Y, Meng Z Y, Zhong C, Lu J, Yu W C, Jia Y B and Qian Y T 2001 Hydrothermal synthesis and characterization of single-molecular-layer $MoS_2$ and $MoSe_2$ *Chem. Lett.* **30** 772-3

[4] Lee Y-H, Zhang X-Q, Zhang W, Chang M-T, Lin C-T, Chang K-D, Yu Y-C, Wang J T-W, Chang C-S, Li L-J and Lin T-W 2012 Synthesis of large-area $MoS_2$ atomic layers with chemical vapor deposition *Adv. Mater.* **24** 2320-5

[5] van der Zande A M, Huang P Y, Chenet D A, Berkelbach T C, You Y, Lee G-H, Heinz T F, Reichman D R, Muller D A and Hone J C 2013 Grains and grain boundaries in highly crystalline monolayer molybdenum disulphide *Nat. Mater.* **12** 554-61

[6] Zhang Y, Chang T-R, Zhou B, Cui Y-T, Yan H, Liu Z, Schmitt F, Lee J, Moore R, Chen Y, Lin H, Jeng H-T, Mo S-K, Hussain Z, Bansil A and Shen Z-X 2014 Direct observation of the transition from indirect to direct bandgap in atomically thin epitaxial $MoSe_2$ *Nat. Nano.* **9** 111-5

[7] Liu H, Jiao L, Yang F, Cai Y, Wu X, Ho W, Gao C, Jia J, Wang N, Fan H, Yao W and Xie M 2014 Dense network of one-dimensional midgap metallic modes in monolayer $MoSe_2$ and their spatial undulations *Phys. Rev. Lett.* **113** 066105

[8] Zhang C, Chen Y, Johnson A, Li M-Y, Huang J-K, Li L-J and Shih C-K 2014 Measuring Critical Point Energies in Transition Metal Dichalcogenides **arXiv:1412.8487v1**

[9] Lehtinen O, Komsa H-P, Pulkin A, Whitwick M B, Chen M-W, Lehnert T, Mohn M J, Yazyev O V, Kis A, Kaiser U and Krasheninnikov A V 2015 Atomic scale microstructure and properties of Se-deficient two-dimensional $MoSe_2$ *ACS Nano* **9** 3274-83

[10] Ugeda M M, Bradley A J, Shi S-F, da Jornada F H, Zhang Y, Qiu D Y, Ruan W, Mo S-K, Hussain Z, Shen Z-X, Wang F, Louie S G and Crommie M F 2014 Giant bandgap renormalization and excitonic effects in a monolayer transition metal dichalcogenide semiconductor *Nat. Mater.* **13** 1091-5

[11] Dean G A 1998 *Langes Chemistry Handbook, 15 Edition* **McGraw-Hill Companies** Section 3

[12] Jiao L *et al* 2015 Molecular-beam epitaxy of monolayer $MoSe_2$: growth characteristics and domain boundary formation *New J. Phys.* **17** 053023

[13] Chiu M-H, Zhang C, Shiu H W, Chuu C-P, Chen C-H, Chang C-Y S, Chen C-H, Chou M-Y, Shih C-K and Li L-J 2014 Determination of band alignment in transition metal dichalcogenides heterojunctions **arXiv:1406.5137v3**

[14] Zhu B, Chen X and Cui X D 2015 Exciton binding energy of monolayer $WS_2$ *Sci. Rep.* **5** 9218

[15] Liu G-B, Xiao D, Yao Y G, Xu X D and Yao W 2015 Electronic structures and

theoretical modelling of two-dimensional group-VIB transition metal dichalcogenides *Chem. Soc. Rev*. **44** 2643-63

[16] Bollinger M V, Lauritsen J V, Jacobsen K W, Nørskov J K, Helveg S and Besenbacher F 2001 One-dimensional metallic edge states in $MoS_2$ *Phys. Rev. Lett.* **87** 196803

[17] Ross J S, Wu S, Yu H, Ghimire N J, Jones A M, Aivazian G, Yan J, Mandrus D G, Xiao D, Yao W and Xu X 2013 Electrical control of neutral and charged excitons in a monolayer semiconductor *Nat. Commun.* **4** 1474

[18] Qiu D Y, da Jornada F H and Louie S G 2013 Optical spectrum of $MoS_2$: many-body effects and diversity of exciton states *Phys. Rev. Lett.* **111** 216805

[19] Cheiwchanchamnangij T and Lambrecht W R L 2012 Quasiparticle band structure calculation of monolayer, bilayer, and bulk $MoS_2$ *Phys. Rev. B* **85** 205302

[20] Ramasubramaniam A 2012 Large excitonic effects in monolayers of molybdenum and tungsten dichalcogenides *Phys. Rev. B* **86** 115409

[21] Komsa H-P and Krasheninnikov A V 2012 Effects of confinement and environment on the electronic structure and exciton binding energy of $MoS_2$ from first principles *Phys. Rev. B* **86** 241201

[22] Ye Z, Cao T, O'Brien K, Zhu H, Yin X, Wang Y, Louie S G and Zhang X 2014 Probing excitonic dark states in single-layer tungsten disulphide *Nature* **513** 214-8

[23] He K, Kumar N, Zhao L, Wang Z, Mak K F, Zhao H and Shan J 2014 Tightly Bound Excitons in Monolayer $WSe_2$ *Phys. Rev. Lett.* **113** 026803

[24] Hanbicki A T, Currie M, Kioseoglou G, Friedman A L and Jonker B T 2015 Measurement of high exciton binding energy in the monolayer transition-metal dichalcogenides $WS_2$ and $WSe_2$ **arXiv:1412.2156**

[25] Wang G, Marie X, Gerber I, Amand T, Lagarde D, Bouet L, Vidal M, Balocchi A and Urbaszek B 2015 Giant enhancement of the optical second-harmonic emission of $WSe_2$ Monolayers by laser excitation at exciton resonances *Phys. Rev. Lett.* **114** 097403

[26] Zhang C, Johnson A, Hsu C-L, Li L-J and Shih C-K 2014 Direct Imaging of Band Profile in Single Layer $MoS_2$ on Graphite: Quasiparticle Energy Gap, Metallic Edge States, and Edge Band Bending *Nano Lett.* **14** 2443-7

[27] Jo S, Ubrig N, Berger H, Kuzmenko A B and Morpurgo A F 2014 Mono- and Bilayer $WS_2$ Light-Emitting Transistors *Nano Lett.* **14** 2019-25

[28] Bordas J 1976 *Some Aspects of modulation spectroscopy in optical and electrical properties* **Springer** 145

[29] Miller R C and Kleinman D A 1985 Excitons in GaAs quantum wells *Journal of Luminescence* **30** 520-40

[30] Ridlay B K 1982 *Quantum Processes in Semiconductors* **Claredon Press, Oxford** P. 209

**Molecular-Beam Epitaxy of Monolayer and Bilayer WSe$_2$: A Scanning Tunneling Microscopy/Spectroscopy Study and Deduction of Exciton Binding Energy**

H J Liu[1], L Jiao[1], L Xie[1], F Yang[2,3], J L Chen[1], W K Ho[1], C L Gao[2,3], J F Jia[2,3], X D Cui[1] and M H Xie[1*]

[1]Physics Department, The University of Hong Kong, Pokfulam Road, Hong Kong

[2]Key Laboratory of Artificial Structures and Quantum Control (Ministry of Education), Department of Physics and Astronomy, Shanghai Jiaotong University, 800 Dongchuan Road, Shanghai 200240, China

[3]Collaborative Innovation Center of Advanced Microstructures, Department of Physics and Astronomy, Shanghai Jiao Tong University, Shanghai 200240, P. R. China

**S1. Determination of the band edges from the STS**

The band edges of ML and BL WSe$_2$ and MoSe$_2$ are determined by plotting the differential conductance (dI/dV) spectra measured by the STS at low temperature in the logarithm scale [1,2]. For this purpose, artificial uniform offsets of the data were made to eliminate negative data due to random errors. Each STS curve shown in the paper represents an average of 50 measurements at the same tip height and the same position far from surface steps and defects. The electronic energy gaps are determined by finding the width of the zero-conductance "floor", $C_{g,av}$, where the band edges are determined by the intersections of the zero-conductance floor with the linear fits of the conductance data in regions of $E_{VB,2\sigma} - \Delta E < E < E_{VB,2\sigma}$ (for valence band edge) and $E_{CB,2\sigma} < E < E_{CB,2\sigma} + \Delta E$ (for conduction band edge) as shown by the red lines in Figure S1. We have chosen $\Delta E = 150$ mV throughout, and so determined band-edges are indicated in the figures shown below.

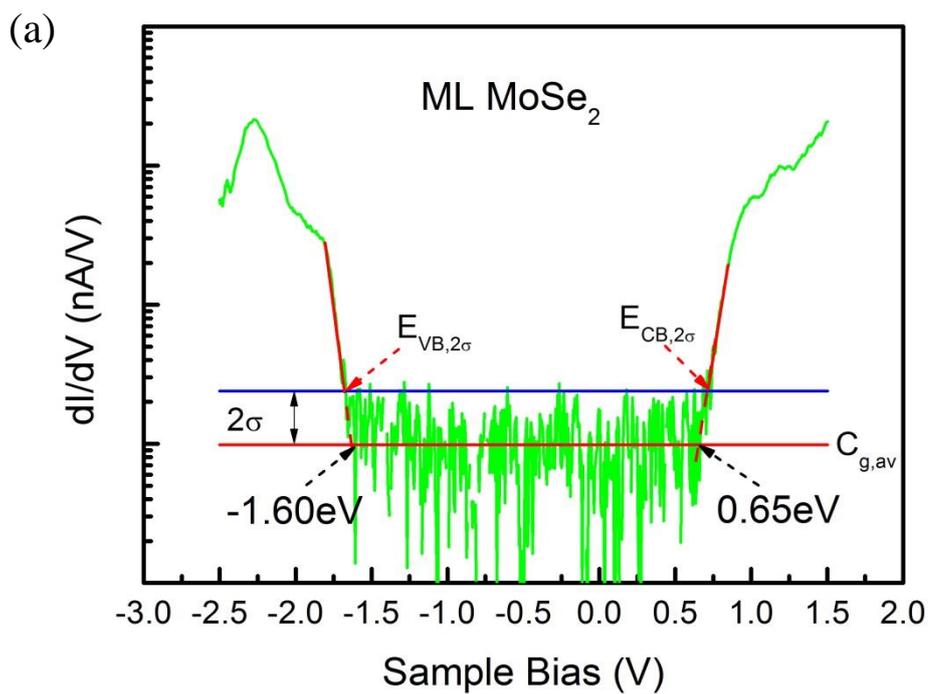

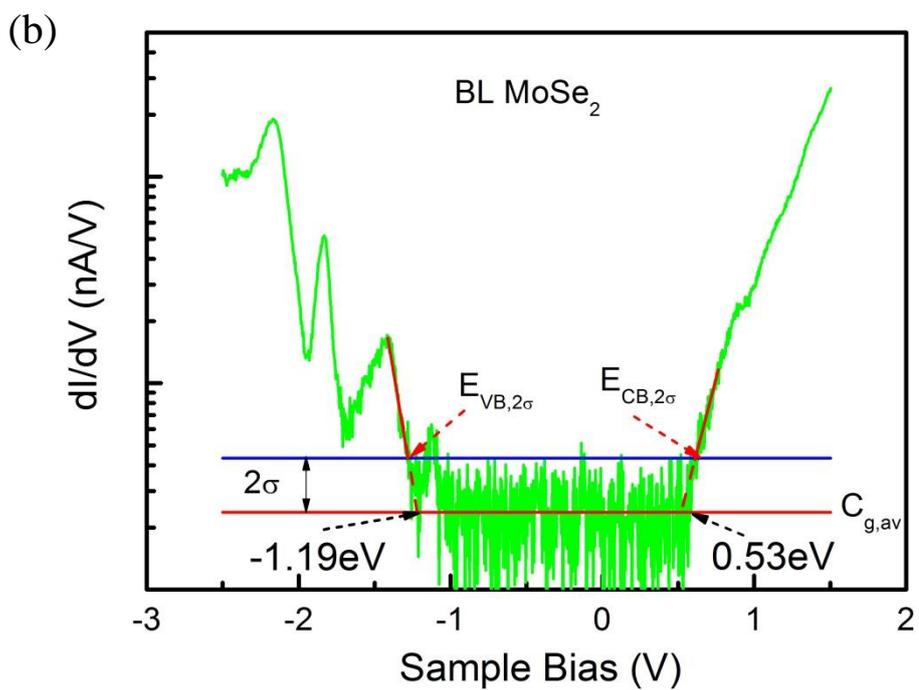

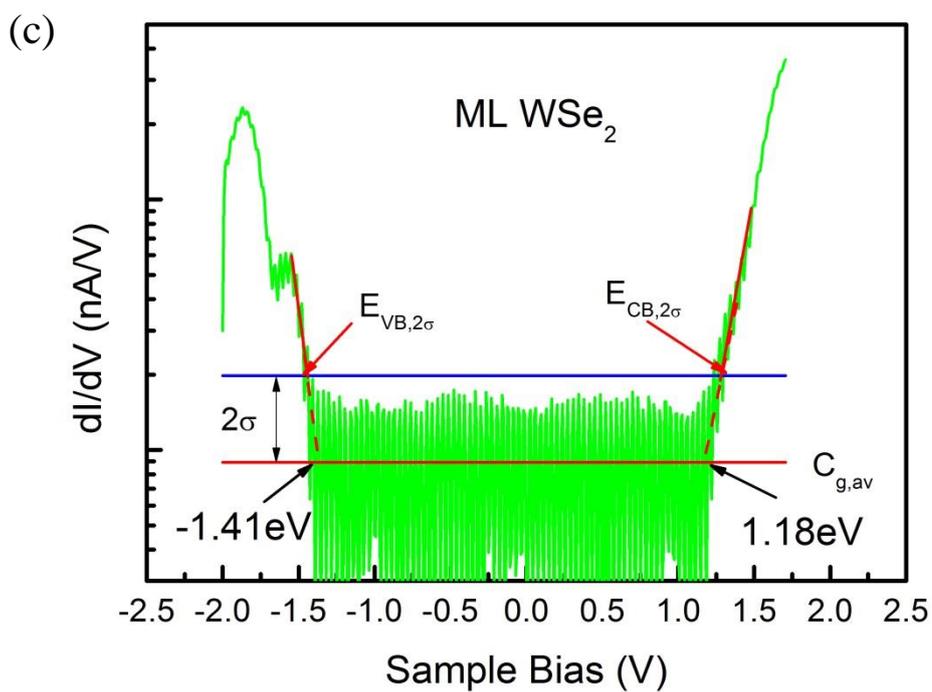

**Figure S1.** Bandgap determination from experimental STS spectra of ML (a, c) and BL (b, d) MoSe$_2$ (a, b) and WSe$_2$ (c, d). The band edges are indicated by dashed arrows.

## S2. PL spectra of BL MoSe$_2$ and WSe$_2$

To evaluate exciton emissions in BL WSe$_2$ and MoSe$_2$, PL spectra were measured at varying temperatures from 10K to 300K in a vacuum cryostat. Figure S2 presents some representative results of the BL samples. In BL MoSe$_2$ (Figure S2(a)), both exciton (X) and trion (X$^-$) emissions are observed at low temperatures and the energy separation of the two gives rise to the binding energy of trions [3,4]. In addition, an intensity shoulder can be noted at the lower energy side of the trion emission in BL MoSe$_2$. The results of multiple peak fittings of the spectra are shown by the dashed green lines in Figure S2(a). The peak (I) embedded in the shoulder is at lower energy and its intensity is much weaker than that of exciton peak (X). The intensity ratio of the peak (I) vs the direct exciton peak (X) becomes dramatically weaker as the temperature decreases and almost invisible at 10K, which is the signature of phonon-assisted transition. Besides, the peak (I) shows a less blue-shift than the direct exciton peak (X) at the decreased temperature. It implies that the states involved in peak (I) are located closer to the Γ point than K point in the Brillouin zone. So we can assign the peak (I) to the band edge of the BL MoSe$_2$, which are at different crystal momenta for the valence and conduction bands.

For BL WSe$_2$, no significant trion emission is detected (Figure S2(b)). Rather a strong and broad emission at 1.5 – 1.6 eV is seen, showing a decreasing intensity with temperature. The origin of such an emission is not very clear, but it is likely related to defects in sample by its temperature dependent behavior. Multi peak fitting of the spectra allows one to determine the energies of the direct (X) and indirect (I) exciton emissions in BL WSe$_2$, which are indicated by the red and black arrows in the figure.

**Figure S2.** PL spectra of BL MoSe$_2$ (a) and WSe$_2$ (b) at different temperatures. Multiple peak fittings of the spectra are shown by the dashed green lines. The energies of the direct and indirect exciton emissions are marked by red and black arrows respectively.

**S3. PL spectrum of ML WSe$_2$ at 9.5 K**

PL measurements of ML WSe$_2$ at 10K reveal unknown peaks, which may again originate from defects in sample. These peaks give way to a single broad peak as the temperature increases to 77K.

**Figure S3.** PL spectra of ML WSe$_2$ measured at 10K (black) and 77K (red), respectively.

**S4. Density Functional Theory (DFT) Calculations**

The DFT calculations are done by the VASP code [5] using the projector augmented wave [6] and Perdew-Burke-Ernzerhof exchange-correlation functional [7]. Relaxed lattice parameters are used for WSe$_2$ monolayer [8]. Spin-orbit coupling is considered in the calculation. The energy cutoff of the plane-wave basis is set to 400 eV and the energy convergence is $10^{-6}$ eV. Vacuum layer is greater than 15 Å to separate neighboring periodic images. A Γ-centered k-mesh of $10 \times 10 \times 1$ is used to obtain the ground-state density and a 51×51 k-mesh is used to calculate the band energies in the rhombus reciprocal cell.

[1] Ugeda M M, Bradley A J, Shi S-F, da Jornada F H, Zhang Y, Qiu D Y, Ruan W, Mo S-K, Hussain Z, Shen Z-X, Wang F, Louie S G and Crommie M F 2014 Giant bandgap renormalization and excitonic effects in a monolayer transition metal dichalcogenide semiconductor *Nat. Mater.* **13** 1091-5

[2] Zhang C, Johnson A, Hsu C-L, Li L-J and Shih C-K 2014 Direct imaging of band profile in single layer MoS2 on graphite: quasiparticle energy gap, metallic edge states, and edge band bending *Nano Lett.* **14** 2443-7

[3] Ross J S, Wu S F, Yu H Y, Ghimire N J, Jones A M, Aivazian G, Yan J Q, Mandrus D G, Xiao D, Yao W and Xu X D 2013 Electrical control of neutral and charged excitons in a monolayer semiconductor *Nat. Commun.* **4** 1474

[4] Zeng H L, Liu G-B, Dai J F, Yan Y J, Zhu B R, He R C, Xie L, Xu S J, Chen X H, Yao W and Xu X D 2013 Optical signature of symmetry variantions and spin-valley coupling in atomically thin tungten dichalcogenides Sci. Rep. 3 1608

[5] Kresse G, Furthmüller J. Efficient iterative schemes for ab initio total-energy calculations using a plane-wave basis set. Phys Rev B 1996, 54(16):

11169-11186.

[6] Blöchl PE. Projector augmented-wave method. Phys Rev B 1994, 50(24): 17953-17979.

[7] Perdew JP, Burke K, Ernzerhof M. Generalized Gradient Approximation Made Simple. Phys Rev Lett 1996, 77(18): 3865-3868.

[8] Liu G-B, Shan W-Y, Yao Y, Yao W, Xiao D. Three-band tight-binding model for monolayers of group-VIB transition metal dichalcogenides. Phys Rev B 2013, 88(8): 085433.